%% file: main.tex
\documentclass[nonacm]{acmart}
\def\BibTeX{{\rm B\kern-.05em{\sc i\kern-.025em b}\kern-.08emT\kern-.1667em\lower.7ex\hbox{E}\kern-.125emX}}
    
\usepackage{nicefrac}
\usepackage{siunitx}
\usepackage{array,framed}
\usepackage{booktabs}
\usepackage{
  color,
  float,
  epsfig,
  wrapfig,
  graphics,
  graphicx,
  subcaption
}

\makeatletter
\renewcommand\@formatdoi[1]{}
\renewcommand\@journalName{}
\renewcommand\@copyrightpermission{}
\def\@copyrightspace{}
\makeatother

\renewcommand\footnotetextcopyrightpermission[1]{}

\usepackage{amssymb}
\usepackage{url}
\usepackage{textcomp,amssymb}
\usepackage{setspace}
\usepackage{latexsym,fancyhdr,url}
\usepackage{enumerate}
\usepackage{algorithm}
\usepackage{algpseudocode}
\usepackage{graphics}
\usepackage{xparse} 
\usepackage{xspace}
\usepackage{multirow}
\usepackage{csvsimple}
\usepackage{balance}
\usepackage{tabularx}

\usepackage{
  tikz,
  pgfplots,
  pgfplotstable
}
\usepackage{hyperref}

\usetikzlibrary{
  shapes.geometric,
  arrows,
  external,
  pgfplots.groupplots,
  matrix
}

\pgfplotsset{compat=1.9}

\usepackage{fancyhdr}
\usepackage{mathtools}

\DeclareMathAlphabet{\mathcal}{OMS}{cmsy}{m}{n}

\DeclareGraphicsExtensions{%
    .png,.PNG,%
    .pdf,.PDF,%
    .jpg,.mps,.jpeg,.jbig2,.jb2,.JPG,.JPEG,.JBIG2,.JB2}

\input{defs}
\makeatletter
\renewcommand\@copyrightpermission{} 
\renewcommand\footnotetextcopyrightpermission[1]{} 
\renewcommand\@mkbibcitation{} 
\makeatother

\setcopyright{none}
\acmConference[]{}{}{}
\acmYear{}
\acmISBN{}
\acmDOI{}
\usepackage{fancyhdr}
\usepackage{graphicx}

\begin{document}

\fancyhead{}

\title[
Evolutionary Defense: Advancing Moving Target Strategies with Bio-Inspired Reinforcement Learning]{Evolutionary Defense: Advancing Moving Target Strategies with Bio-Inspired Reinforcement Learning to Secure Misconfigured Software Applications
}

\author{
    Niloofar Heidarikohol\textsuperscript{\rm 1},
    Shuvalaxmi Dass\textsuperscript{\rm 1},
    Akbar Siami Namin\textsuperscript{\rm 2} \\
}
\affiliation{
    \textsuperscript{\rm 1} School of Computing and Informatics, University of Louisiana at Lafayette\\
    \textsuperscript{\rm 2} Department of Computer Science, Texas Tech University\\
    niloofar.heidarikohol1@louisiana.edu; shuvalaxmi.dass@lousiana.edu; akbar.namin@ttu.edu
}


\input{abstract}
\maketitle
\thispagestyle{plain}


\input{1intro}    
\input{2Prelim}

\input{3RelatedWork}

\input{4background}
\input{5MTDAlgorithmDesign}

\input{6ExperimentsandResults}
\input{7Discussion}

\input{9LiteratureReview}
\input{8ConclusionandFuturework}

\bibliographystyle{plain} 

\bibliography{bib}

\appendix
\input{appendix}
\end{document}

%% file: defs.tex
\usepackage{xparse}
\newcommand{\bnm}{\begin{newmath}}
\newcommand{\enm}{\end{newmath}}

\newcommand{\bea}{\begin{eqnarray*}}%
\newcommand{\eea}{\end{eqnarray*}}%

\newcommand{\bne}{\begin{newequation}}
\newcommand{\ene}{\end{newequation}}

\newcommand{\bal}{\begin{newalign}}
\newcommand{\eal}{\end{newalign}}

\newenvironment{newalign}{\begin{align}%
\setlength{\abovedisplayskip}{4pt}%
\setlength{\belowdisplayskip}{4pt}%
\setlength{\abovedisplayshortskip}{6pt}%
\setlength{\belowdisplayshortskip}{6pt} }{\end{align}}

\newenvironment{newmath}{\begin{displaymath}%
\setlength{\abovedisplayskip}{4pt}%
\setlength{\belowdisplayskip}{4pt}%
\setlength{\abovedisplayshortskip}{6pt}%
\setlength{\belowdisplayshortskip}{6pt} }{\end{displaymath}}

\newenvironment{newequation}{\begin{equation}%
\setlength{\abovedisplayskip}{4pt}%
\setlength{\belowdisplayskip}{4pt}%
\setlength{\abovedisplayshortskip}{6pt}%
\setlength{\belowdisplayshortskip}{6pt} }{\end{equation}}

\newcounter{ctr}

\newcounter{mytable}
\def\mytable{\begin{centering}\refstepcounter{mytable}}
\def\endmytable{\end{centering}}

\newcounter{myfig}
\def\myfig{\begin{centering}\refstepcounter{myfig}}
\def\endmyfig{\end{centering}}

\newlength{\saveparindent}
\newlength{\saveparskip}
\newcommand{\E}{{\rm I\kern-.3em E}}

\renewcommand{\eqref}[1]{\mbox{Equation~(\ref{#1})}}

\def \part {part}

\renewcommand{\paragraph}[1]{\vspace*{6pt}\noindent\textbf{#1}\;}

\def \blackslug{\hbox{\hskip 1pt \vrule width 4pt height 8pt
    depth 1.5pt \hskip 1pt}}
\def \qed{\quad\blackslug\lower 8.5pt\null\par}

\newcounter{mynote}[section]

\newcommand\ignore[1]{}

\newcounter{rcnote}[section]

\newcounter{mrnote}[section]

\newcounter{fknote}[section]

\newcounter{anote}[section]

\DeclareMathSymbol{\mlq}{\mathord}{operators}{``}
\DeclareMathSymbol{\mrq}{\mathord}{operators}{`'}

\newcommand{\rhf}[2]{R_{f, \gamma}}

\DeclareDocumentCommand{\edist}{o o}{
  \ensuremath{
    \IfNoValueTF{#1}{{d}}{{\sf d}(#1,#2)}
  }
}

\newcommand{\olrk}[1]{\ifx\nursymbol#1\else\!\!\mskip4.5mu plus 0.5mu\left(\mskip0.5mu plus0.5mu #1\mskip1.5mu plus0.5mu \right)\fi}

\NewDocumentCommand{\indseq}{ O{1} O{r} }{{#1}\ldots {#2}}


%% file: abstract.tex
\begin{abstract}
\textbf{Abstract:} Improper configurations in software systems often create vulnerabilities, leaving them open to exploitation. Static architectures exacerbate this issue by allowing misconfigurations to persist, providing adversaries with opportunities to exploit them during attacks. To address this challenge, a dynamic proactive defense strategy known as Moving Target Defense (MTD) can be applied. MTD continually changes the attack surface of the system, thwarting potential threats. In the previous research, we developed a proof of concept for a single-player MTD game model called RL-MTD, which utilizes Reinforcement Learning (RL) to generate dynamic secure configurations. While the model exhibited satisfactory performance in generating secure configurations, it grappled with an unoptimized and sparse search space, leading to performance issues. To tackle this obstacle, this paper addresses the search space optimization problem by leveraging two bio-inspired search algorithms: Genetic Algorithm (GA) and Particle Swarm Optimization (PSO). Additionally, we extend our base RL-MTD model by integrating these algorithms, resulting in the creation of PSO-RL and GA-RL. We compare the performance of three models: base RL-MTD, GA-RL, and PSO-RL, across four misconfigured SUTs in terms of generating the most secure configuration. Results show that the optimal search space derived from both GA-RL and PSO-RL significantly enhances the performance of the base RL-MTD model compared to the version without optimized search space. While both GA-RL and PSO-RL demonstrate effective search capabilities, PSO-RL slightly outperforms GA-RL for most SUTs. Overall, both algorithms excel in seeking an optimal search space which in turn improves the performance of the model in generating optimal secure configuration.
\\
\textbf{\textit{Keywords:}} Software engineering, Software configuration management, Moving Target Defense, OS security

\end{abstract}





%% file: 1intro.tex
\section{Introduction}
\label{sec:intro}
In today's digital era, software applications are integral to every aspect of our lives, from powering our smartphones to driving complex business operations, driving innovation and efficiency. However, such highly configurable applications if not configured properly can lead to security misconfiguration in software making the system vulnerable to attacks like data breaches. In 2021, Twitch, an interactive live-streaming platform, suffered a massive 125GB data and source code leak due to server misconfiguration.\cite{t}\newline
The Open Web Application Security Project (OWASP) lists "Security misconfiguration" among the top 5 web application security risks as of 2021\cite{ow}. Misconfigurations stem from the \textit{static nature} of application configurations, persisting over time. This static characteristic can result in improper security settings due to missed updates, human errors, or incomplete configurations, leaving systems vulnerable to attacks or unauthorized access by attackers\cite{why}.
To counter such threats, we require a \textit{dynamic} defensive strategy to overcome the static nature of configurations. Traditional solutions like costly antivirus software, which focuses on detection and reaction, are ineffective against evolving attack strategies. A proactive dynamic defense is a more effective approach. \newline 
In our preliminary work, we introduced a model for generating dynamic secure configurations using \textbf{Moving Target Defense (MTD)} as our proactive defense solution. We implemented MTD via Reinforcement Learning (RL), termed RL-MTD. However, the performance of our RL-MTD base model was impacted by sparse search space issues. In this paper, we address the performance issue of non-optimized search space by enhancing our base model through integration with bio-inspired algorithms (GA-RL and PSO-RL) to improve performance. \newline
\textbf{MTD:} According to the Department of Homeland Security (DHS),  MTD is a military strategy translated to the cybersecurity world that involves dynamically manipulating various system configurations to alter and manage the attack surface, thereby increasing uncertainty and complexity for attackers \cite{dod}. This approach reduces opportunities for attackers to identify vulnerable system components and raises the cost of launching attacks or scans. Ultimately, the goal is to make attackers expend time and effort without gaining valuable intelligence about the system.\cite{survey}. \newline
To the best of our knowledge, our work represents the initial endeavour to introduce a proof-of-concept MTD defense strategy focused on software configuration at the individual application level. This paper builds upon preliminary research that laid the groundwork for developing the RL-MTD framework to address software misconfiguration as follows:
\begin{enumerate}
    \item We briefly discuss building our foundation work which is base RL-MTD framework.
    \item The optimization issue of the search space in the RL-MTD model is examined.
    \item We propose the integration of RL-MTD with bio-inspired algorithms (GA, PSO) to create GA-RL and PSO-RL, offering a solution for optimizing the search space problem.
    \item A quantitative analysis is conducted to compare the performance of RL-MTD, GA-RL, and PSO-RL across four misconfigured case studies of (SUTs) in terms of generating secure configurations.
\end{enumerate}

The paper follows this structure: an overview of prior work on RL-MTD model (Section 2), addressing the search space issue (Section 3), integrating PSO and GA to RL-MTD (Section 5) after providing background on PSO and GA (Section 4), experimental setup and results (Section 6), discussion of results, related work, and conclusion with future work (Sections 7-9).

%% file: 2Prelim.tex
\section{preliminary Work}
\label{sec:relwork}
In this section, we will first briefly discuss our preliminary work which forms the foundation of the initial work done in the field of Moving Target Defense for software misconfiguration.
We will divide this section into 2 parts where the first part talks briefly about the motivating example/problem statement and the second part demonstrates the MTD approach designed for the problem.

\subsection{Motivating Scenario}
Consider a host machine in an organization with a misconfigured software system, vulnerable to attacks like brute force/credential stuffing, code injection, buffer overflow, and cross-site scripting (XSS). Attackers exploit these vulnerabilities by first conducting reconnaissance to understand the misconfigurations. To counter this, the proposed solution uses a Moving Target Defense (MTD) strategy that frequently changes the configuration settings. This approach aims to create a dynamic, secure environment, rendering the attacker's knowledge obsolete and preventing effective exploit development \cite{jalowski2022survey}. \textbf{The goal is to move towards secure configurations via dynamically changing the configurations from misconfigured state to a secure state as a dynamic defensive measure (MTD) using Reinforcement Learning (RL)}. This will confuse the attackers who rely on outdated or constantly evolving information.

\subsection{RL-MTD Approach}
The idea is to develop a proof-of-concept of a defense approach inspired by the Moving Target Defense (MTD) to defend misconfigured SUT. We modeled MTD in the form of a single-player game implemented using the model-free Monte-Carlo method in Reinforcement Learning (RL) where the goal is to convert a misconfigured SUT to a secure-configured SUT as a means to defend it from potential attacks. This section first talks about the attack surface 
used in our MTD strategy followed by the game model description using RL.

\subsubsection{Attack Surface}
\begin{figure}[h]
 \centering
      \includegraphics[width=6cm]{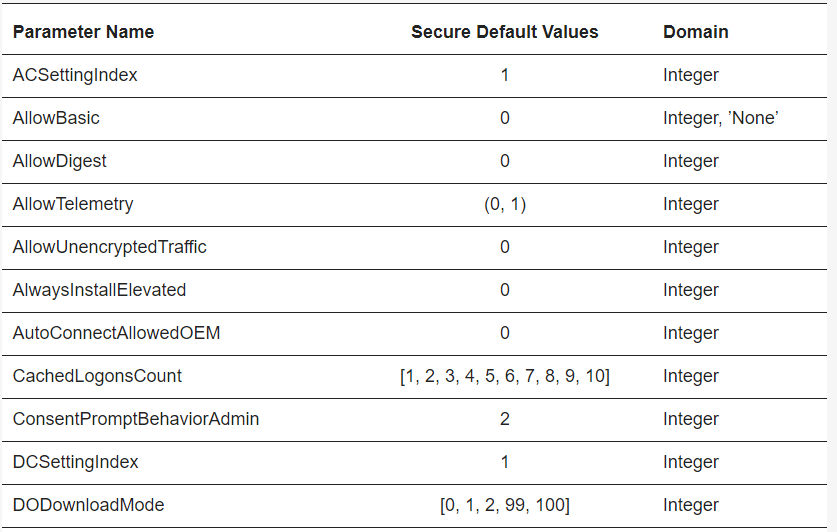}
      \caption{A subset of Windows 10 configuration parameters associated with default secure settings}
      \label{fig:STIG}
\end{figure}

We represent the attack surface of a SUT as a configuration $C$ which is composed of a series of parameters P that belong to SUT. We denote $C$ as: \newline
            \centerline {$C:= \{ P_1:S_1, P_2:S_2, \ldots, P_n:S_n \}$ }
where $n$ represent the number of parameters and $S_i$ is the setting value associated with parameter $P_i$ in a configuration space of SUT. We collected the configuration information of different SUT from the Security Technical Implementation Guide (STIG) \cite{stig}. The STIG guidelines offer proper checklists to view the “compliance status” of the system’s security settings. In other words, the STIG checklists enable us to test whether the underlying system configuration complies with standards (i.e., secure system settings regulations. Figure \ref{fig:STIG} lists some of the parameters of Windows 10 along with their default values and the domain of values it belongs to. However, the key goal of the MTD technique \cite{survey}  is to rearrange or randomize system configurations to increase confusion and uncertainty for attackers. Therefore, the attack surface that we use for the RL-MTD model, starts with a misconfigured SUT  and the task of the agent is to learn too navigate towards the securely configured SUT.
A misconfigured attack surface would have all the parameter's settings improperly set (i.e randomly drawn from  domain of (P))
For instance, C for misconfigured Windows would look like: \newline
 \centerline { $C:= \{ ACSettingIndex: 5, AllowBasic:3, \ldots, DoDownloadMode:4 \}$ }
where the settings are a finding as these parameters are not securely set. 

\subsubsection{RL-Based Game Description of MTD}
\begin{figure}[htbp]
\centering
\includegraphics[width=0.9\linewidth]{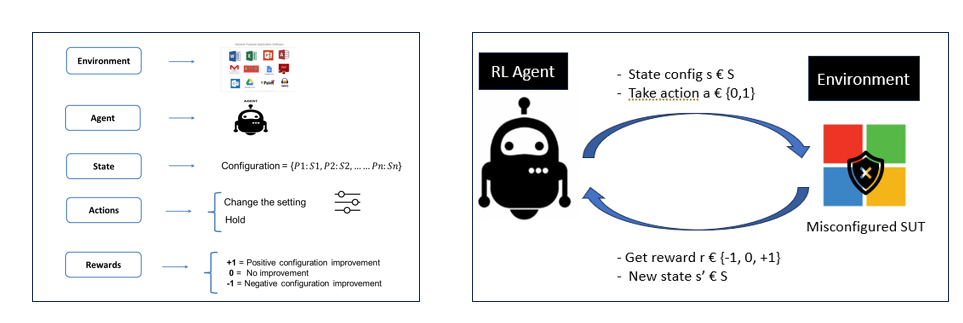}
      \caption{\textbf{(Left)} RL elements: The environment is the misconfigured SUT's attack surface, an agent is the Monte-Carlo-based RL agent, the State represents the configuration (C) instance/state of the misconfigured SUT, actions taken by an agent are to either change a particular parameter setting or hold back and rewards are given based on the improvement(+,- or none) of configuration security score from its previous state. \textbf{(Right)} The MTD-RL agent interacts with a misconfigured SUT environment(eg Windows), where the state s is the current config (C) it is in, and it takes an action (0 or 1) based on which the SUT moves the agent to the next state s' and returns a  reward (0,1,-1) based on the actions }
      \label{fig:RLel}
\end{figure}

MTD is modeled as a \textit{single-player game} played by an RL agent which acts as a defender focusing on the security of misconfigured system. Imagine SUT's attack surface as a board game where the MTD RL agent has a start state, a goal state, and multiple dynamic intermediate states which are dynamically generated and traversed based on the RL agent's action. The overall objective of the MTD-RL agent is to dynamically transition from an insecure state (start) of the misconfigured SUT  towards a nearly secure state(goal) of the SUT by taking some actions. Figure \ref{fig:RLel} shows the RL environment elements used in our MTD approach and how RL elements interact with the environment which is the attack surface of a misconfigured SUT. 

\begin{figure}[htbp]
\begin{center}
      \includegraphics[width=0.9\linewidth]{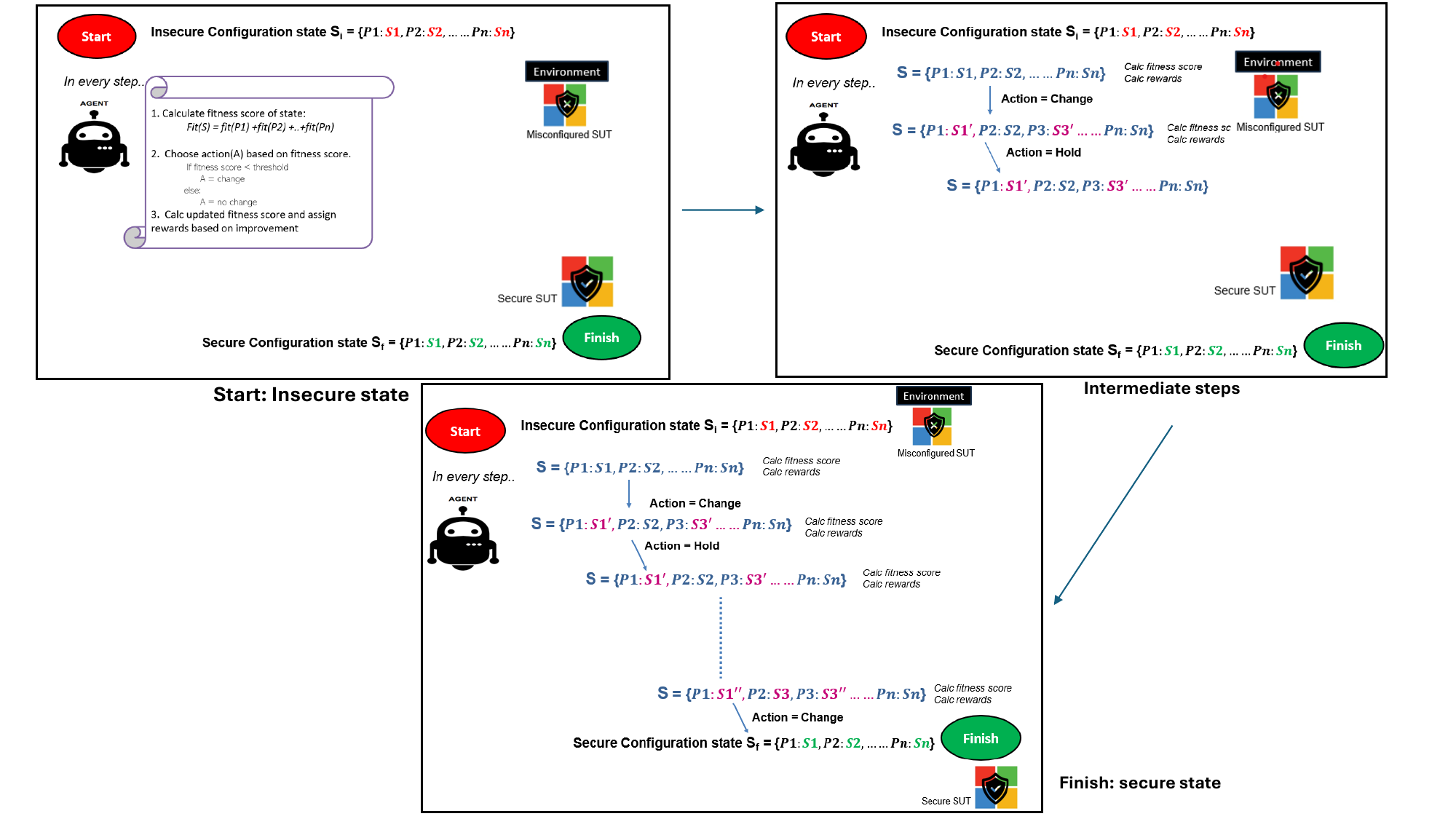}
      \caption{This shows the snapshot of what an RL-MTD game looks like when played. Initially, it starts from an insecure state (\textbf{Top-left}), takes an action based on the config fitness score, gets a reward, and moves to the next intermediate steps (\textbf{top-right}). This series of operations is followed in every step until the agent reaches the near-optimal secure config finish state (\textbf{bottom}). }
      \label{fig:game}
      \end{center}
\end{figure}

Starting from the initial state, at every step, the agent must decide whether to alter (action = change (1)) or keep some parameter settings as is (action = Hold(0)) of the current config state. The action taken is based on the fitness score of the current config state, where if it is below a certain threshold value, action  = 1 is chosen, other 0. Subsequently, a reward is generated which indicates how good the action is which helps the maximize its choices to progressively achieve a more secure intermediate next config state until it reaches the goal state. Figure \ref{fig:game} shows how the game is played where the ultimate strategy is to continuously and dynamically modify the attack surface towards the direction of the finish/secure state, thereby reducing/changing vulnerabilities in the process and confusing the attackers by increasing the uncertainty. The game's dynamics are implemented through the model-free RL-based Monte Carlo Prediction method, which guides the decision-making process for optimizing security configurations through a defined policy.

\subsubsection{RL-MTD Game Algorithm}
This section outlines a procedure and a series of step-by-step algorithms for implementing the environment required to execute model-free RL Monte Carlo (MC) prediction method in the context of the RL-MTD game. The Monte Carlo method can help improve a supplied policy that is effective at making decisions that lead to winning the game which in our case is reaching as close to the terminal state as possible. In other words, the purpose of this method is to generate secure configurations for a misconfigured SUT platform by assessing the quality of a given policy.
Technically, in MC prediction, the objective is to estimate the state-value function V(s), which represents the expected return from a state S under a given policy. Instead of using the term "expected return" (which is the discounted sum of rewards), we employ the concept of "empirical return." Essentially, MC prediction assesses how well a fixed predefined policy performs by predicting the mean total rewards from any given state, assuming the policy remains constant \cite{RLBook}. The MC prediction pseudocode is shown in Algorithm 1 \newline
Our RL- MTD algorithm is also supplied with a fixed policy and aims to evaluate its performance in terms of the value function. In other words, our goal is to predict the expected total reward from the most secured state it as reached. Consequently, we measure the reward using the fitness score of each model configuration, where higher fitness indicates a more secure configuration. Moreover, we chose a model-free MC prediction method as the probability of transitioning (i.e., transition probabilities) to the next configuration/state (different set of settings) cannot be gauged from the environment as there is no pre-defined domain of knowledge known to measure the likelihood of moving from one configuration to another. As a result, the agent learns through running multiple episodes, constantly collecting samples (random values of settings), getting rewards, and thereby evaluating the value function.
The RL-MTD approach explained in the aforementioned subsection can be divided into the following steps:
\begin{enumerate}
    \item \textbf{Step 1: Set Initial State.} Initial state is a configuration C of the underlying system initialized with random settings for its parameters.
    \item \textbf{Step 2: Compute config Score.} The fitness score of a configuration state is the total sum of individual fitness scores of the parameters. A parameter receives a definite $HIGH = 800$ score if it is associated with its secure setting according to the STIG website. Otherwise, a $LOW =8$ score is assigned. The fitness score indicates how secure a configuration is. The higher the fitness score, the more secure it is. These are hyperparameter values used to score the severity of these settings.
    \item \textbf{Step 3: Set Action Policy.} If the overall fitness score of the configuration is below a certain $threshold$ value, then the agent chooses action \textit{a} either 0(hold) or 1(change) based on the  probability distribution \textit{p(a)} as follows:
         \begin{center}
$p(a) = 
\begin{cases}
    \{prob_{0} = low, prob_{1} = high\}: \& \\ \text{if \textit{$ fitness\_score < Threshold $}} \\
    \{prob_{0} = high, prob_{1} = low\}: \& \\ \text{if \textit{$ fitness\_score >= Threshold $}} \\
\end{cases}
$
\end{center}
where $a = hold(0) \ or \ Change(1)$. This ensures the agent chooses action 1 more if the fitness score of the C is not up to the mark (threshold) and vice-versa
   \textbf{ \item Step 4: Generate <S,A,R> tuple.} This tuple is indicative of the current position of the agent in the attack surface. Rewards measure how good of an action was taken by calculating the fitness improvement (Fit(new S) - Fit(old S))
   \textbf{\item Step 5: Generate Episodes.} Collection of tuples used for training the agent.
    \textbf{\item Step 6: Execute RL MC prediction with the given action policy.}
    This is executed with multiple episodes and eventually captures the best fitness scores in the form of the value of state V which indicates how secure a given state is:
    \begin{equation}
V_s = E_{\pi}[R_{t+1} + \gamma R_{t+2} + \gamma^2 R_{t+3}...| S_t=s] 
\end{equation}

where $E$ is the expected mean of the reward for the state $s$.
\end{enumerate}

As there was no pre-existing environment for our problem domain in OpenAI gym \cite{open} at our disposal, we had to create our environment using Step 4  which required Steps 1, 2, and 3.
Step 5 is used to generate an episode of 100 <S,A,R> tuples that are used for training the MC prediction algorithm as shown in Algorithm \ref{alg:MCP} which is a standard algorithm used in literature. \newline
In short, RL-MTD model is composed of 3 main parts: environment(), generate\_episode() and MC\_prediction method ()
  
\begin{algorithm}[!h]
    \caption{Monte Carlo Prediction Pseudocode.(Excerpt taken from \cite{RLBook}) }\label{alg:MCP}
    \begin{algorithmic}[1]
        \Procedure{mc\_prediction}{$policy, num\_ep, df $}
        \State $returns\_sum \gets \{  \}$ \Comment{Keeps track of sum of returns for each state
        to calculate an average.}
        \State $returns\_count \gets \{  \}$ \Comment{Keeps track of count  returns for each state to calculate an average.}
        \State $V \gets \{  \}$ \Comment{The final value function}
        \For {$i$ in $range(1, num\_ep+1)$}
            \State $episode \gets generate\_episode(policy)$
            \State $states\_in\_episodes \gets$ Find all states visited in this episode and convert them into tuple
            \For {$state$ in $states\_in\_episodes$}
                \State $first\_occurrence \gets$ First occurrence of the state in the episode
                \State $G \gets$ Sum up all rewards since the first occurrence
                \State $returns\_sum[state] += G$
                \State $returns\_count[state] += df$
                \State $V[state]$ = $returns\_sum[state]/returns\_count[state]$
            \EndFor
        \EndFor\label{MCPFor}
        \State \textbf{return} $V$
    \EndProcedure
    \end{algorithmic}
\end{algorithm}

\begin{figure}[h]
 \begin{center}
      \includegraphics[width=0.9\linewidth]{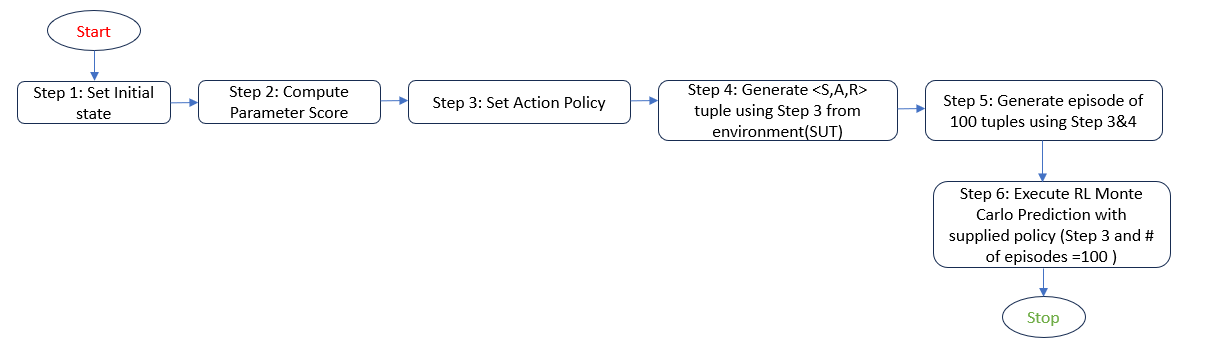}
      \caption{The execution flow of base RL-MTD model where Steps 1-4 make up the environment(), step 5 is generate episode() and step 6 is MC\_prediction method ()}
      \label{fig:floRL}
    \end{center}
\end{figure}

Figure \ref{fig:floRL} shows the complete execution process flow of our RL-MTD algorithm

%% file: 3RelatedWork.tex
\section{Search Space issue in RL-MTD}
\begin{figure}[h]
 \begin{center}
      \includegraphics[width=9cm]{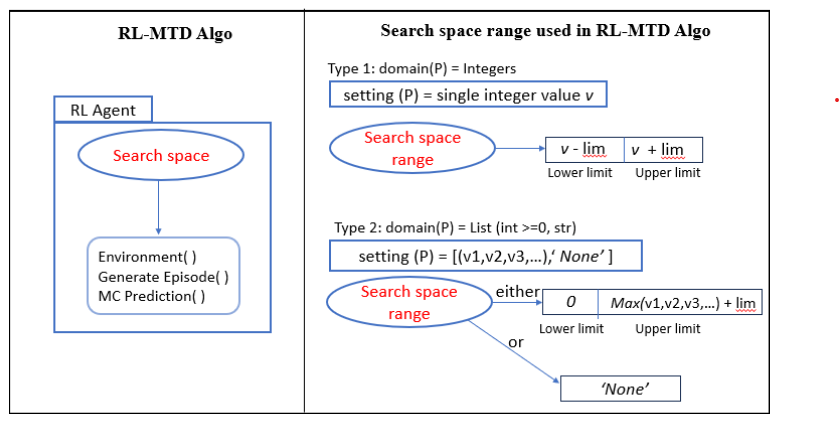}
      \caption{\textbf{Left:} RL-MTD Algo shows the diagrammatic view of all the important functions that use the search space/domain corresponding to a particular SUT. This search range is used by the agent to randomly draw settings from either during the initial config state or when action=1 is chosen.
      \textbf{Right:} shows the search space range for 2 types of parameter settings where the agent has to pick a setting from \{v-lim,v+lim\} if default setting(P) is a single integer value \textit{v} and \textit{lim} is a hyperparameter for a limit of int type; and if default setting(P) could be any value from a list consisting of permissible non-neg integers (v1,v2,v3,...) and/or 'None', it gets to choose either a numerical val from \{0, max( (v1.v2,v3,.)+lim\} or the string value 'None'.
      \textbf{Issue:} How to effectively choose value of \textit{lim} that will optimize the search space range for better performance?}
      \label{fig:search}
      \end{center}
\end{figure}

In the RL-MTD work, there are two instances where the RL agent has to draw random setting values from the search space range of every parameter $P$ present in a configuration: once during the start state (initial state) of the game when it has generated a random configuration state, other times whenever the action = 1 is chosen and it has to change settings of low score parameters.\newline
As per the STIG website, the permissible default settings are defined for any parameter $P$ belonging to a particular SUT. In our work, as this is a proof-of-concept, we handpicked only those parameters that mostly had integer values and/or 'None' as their settings for ease of computation.
Figure \ref{fig:search} shows how we designed the search space range for the agent to choose from during the RL-MTD operation for any particular SUT. The agent has to learn to eventually choose the permissible secure setting for every parameter P in configuration C from the given search range to ensure  C is close to being secure (goal state). We set the custom space range as follows depending on the data type of setting(P):

\begin{enumerate}
    \item Numeric: $(v-lim, v+lim)$, and
    \item List: choice between $[(0, max(v1,v2,v3..) +lim]$ and  $None$
\end{enumerate}

where $v$ is the default setting value as per the STIG website for any particular SUT, $max(List)$ is the maximum setting value if it is a list type, and $lim$ is an arbitrary integer value. We set $lim = 10$ based on multiple experiments we conducted and this value seemed to give better secure configurations (as we also covered in the Results section).\\
\textbf{Issue:} However, finding the right value of \textit{lim} to define the search space range from where the RL agent picks up random values for parameters can be a time-consuming task and often requires us to play around with a bunch of different values before we can find a good enough candidate. As this search space range is majorly responsible for the performance of RL in terms of generating secure configurations, there is a need to find an optimized search space for RL that is best for agents to generate diverse yet secure configurations. \newline
The solution to this problem is to use search optimization algorithms. In section 5,  we describe how we integrate the bio-inspired search optimization algorithms namely Genetic Algorithm(GA) and Particle Swarm Optimization(PSO) into RL-MTD  and develop Evolutionary RL(E-RL) algorithms for better performance.\newline
We integrate both GA and PSO in our approach because they are widely recognized optimization techniques. PSO operates as a population-based stochastic optimization algorithm, focusing on the collective behavior of swarms \cite {psoadvantage}, while GA functions as a heuristic search-based algorithm, simulating evolutionary processes like crossover and mutation \cite{GA}. Our study aims to compare and assess the performance of these two algorithms to determine which one surpasses the base model in effectiveness.

%% file: 4background.tex
\section{Background}
\label{sec:bck}
In this section, we briefly discuss the general working of bio-inspired algorithms: Genetic Algorithm and Particle Swarm optimization
\subsection{Genetic Algorithm}
\begin{figure}[h]
 \centering
      \includegraphics[width=5cm]{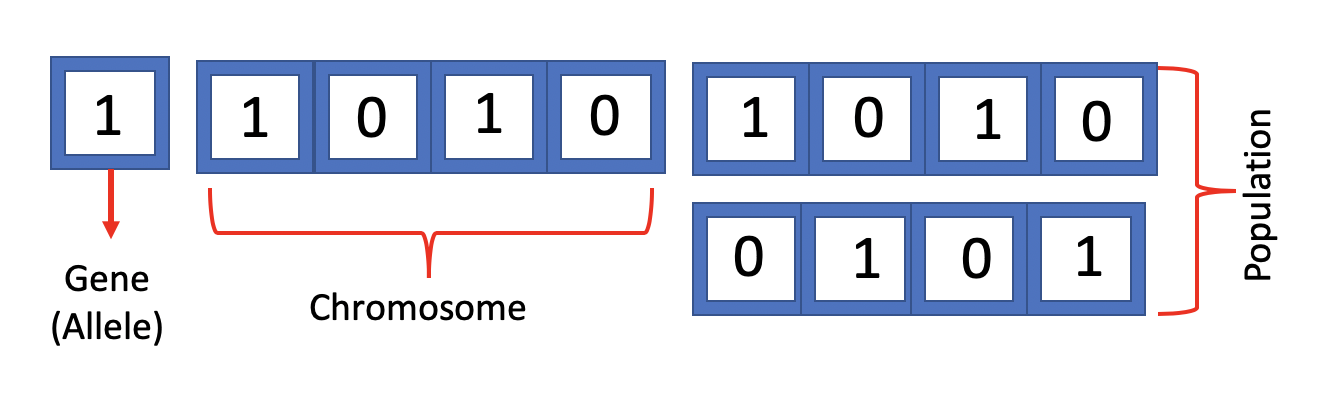}
      \caption{The elements of genetic algorithm.}
      \label{fig:GAelements}
\end{figure}

\begin{figure}[h]
 \centering
      \includegraphics[width=5cm]{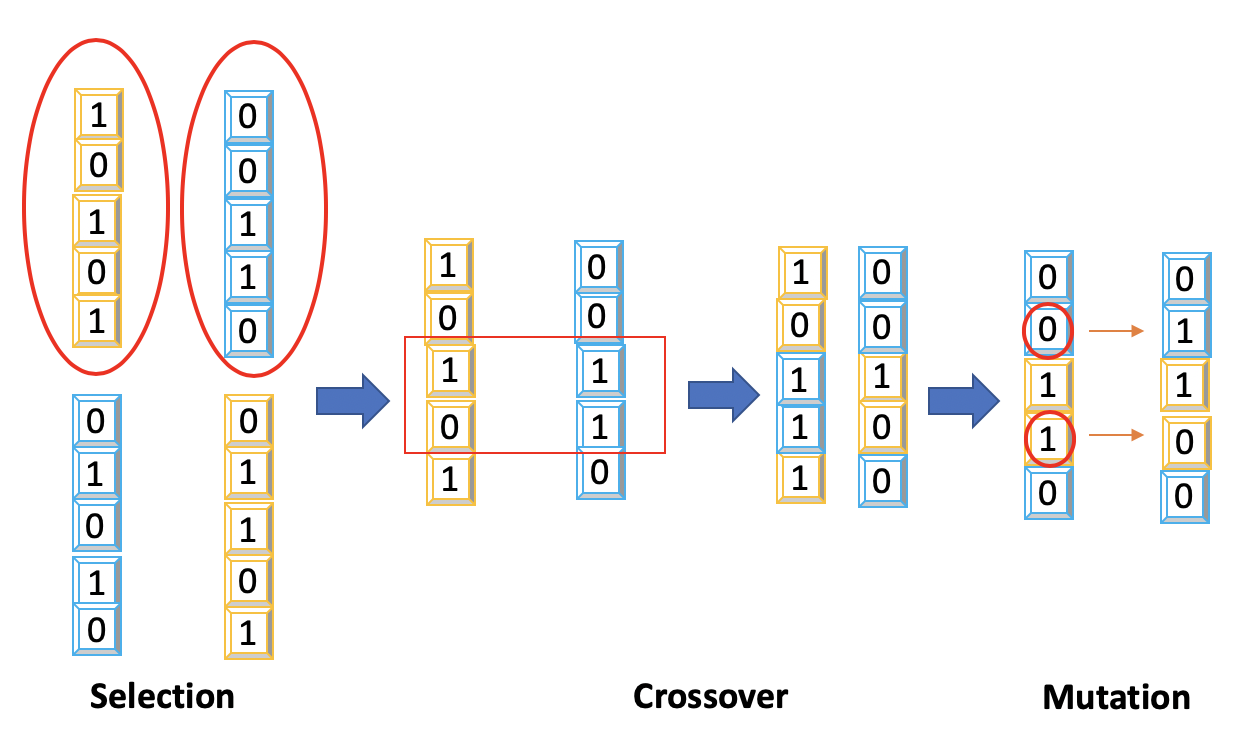}
      \caption{Genetic algorithm process.}
      \label{fig:GAprocess}
\end{figure}

Genetic Algorithms(GA) are based on the biological process of evolution. The idea is that over time, a pool of chromosomes will evolve to be even better (i.e., better fitness value) than the previous generation. A new generation (equal to the pool size) of chromosomes (i.e., configurations) is created with any iteration of the algorithm. This is achieved by the processes of selection, crossover, and mutation \cite{ga1}. A fitness score metric is adopted as a measure to select the two fittest chromosomes from the pool that are called parent chromosomes. Then crossover takes place between the parents to produce a new child chromosome, which will have the best traits from both the parents followed by mutating of some of the characteristics of the child to introduce new traits. This process is repeated until an entirely new generation gets created. Figures \ref{fig:GAelements} and \ref{fig:GAprocess} show the elements and the process of GA respectively. 

\subsection{Particle Swarm Optimiziation}  

The Particle Swarm Optimization approach is based on natural bio-inspired systems, including bird flocks or schooling fish~\cite{huang2023optimized}.
Each particle follows certain fundamental rules to navigate the search space effectively or reach optimal values~\cite{song2004research}.
Individuals should maintain an appropriate distance in standard scenarios, avoid collisions, and remain very close when confronted with threats~\cite{huang2023optimized}.  

\begin{figure}[h]
 \centering
      \includegraphics[width=5cm]{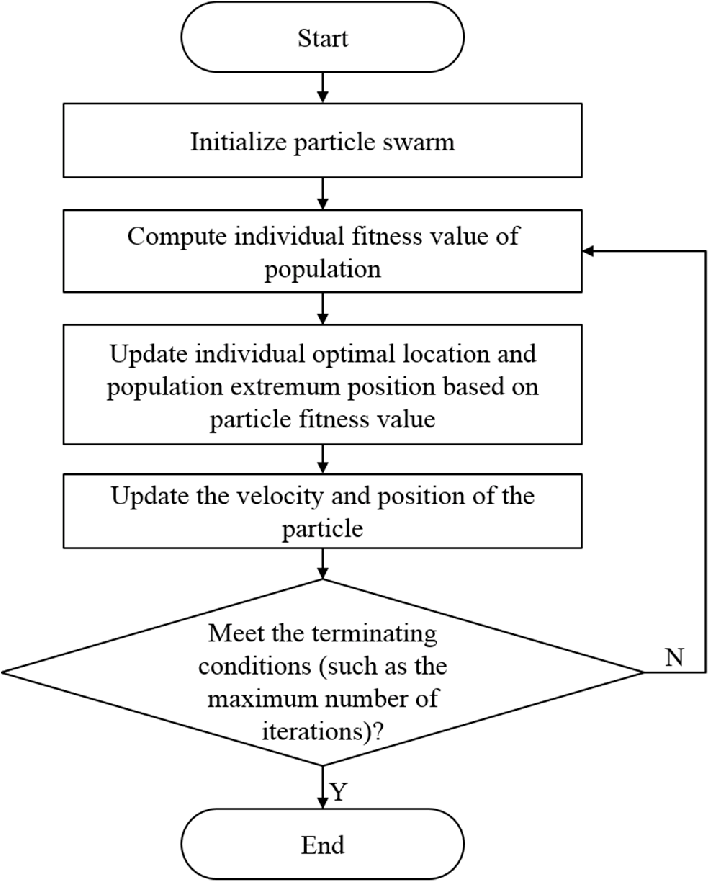}
      \caption{Standard PSO flowchart process.}
      \label{fig:PSOflow}
\end{figure}

PSO stands as an iterative, random, and population-based optimization algorithm for determining the optimal value.
This is accomplished by assigning a particle to locate the ideal location or answer within the search space.
Each particle's dynamics is influenced by social movements as well as its own internal dynamic.
Originally, each particle, irrespective of its peers, can be considered to behave independently to ameliorate its behavior.
However, swarms attempt to adjust to the behavior of other particles as the algorithm processes.
Consequently, each particle adjusts updates iteratively with other particles to observe the optimal value.
Furthermore, the characteristics of each swarm can be determined by the interaction of position and velocity~\cite{papazoglou2023review}. 
Figure \ref{fig:PSOflow} shows the workflow of a general PSO process\cite{psoflow}. The algorithm starts by initializing a swarm of particles with random positions and velocities. It then enters a loop where it evaluates the fitness of each particle, updates their personal best positions and the global best position, and then adjusts their velocities and positions accordingly. This loop continues until a termination criterion, such as a maximum number of iterations or a satisfactory fitness level, is met.

%% file: 5MTDAlgorithmDesign.tex
\section{Bio-inspired RL-MTD Modelling}
\label{sec:methodology}
In this section, we delve into the design of the RL-MTD strategy through the an integrated approach using a Genetic Algorithm with Reinforcement Learning(GA-RL) and Particle Swarm Optimization with RL (PSO-RL). We will elucidate how the components of each algorithm - GA-RL, and PSO-RL - are adapted to our specific context of generating an optimized search space for our RL-MTD algorithm to perform better in generating secure configuration. This is followed by a detailed flowchart presentation, illustrating how these individual strategies are operationalized.
\subsection{Elements Representation for GA-RL, PSO-RL}
Before we dive into the algorithm flow of how MTD is realized using each of these methods, we will first show the element representation used for each bio-inspired algorithm.

\subsubsection{GA-RL} 
\begin{figure}[h]
 \centering
      \includegraphics[width=8cm]{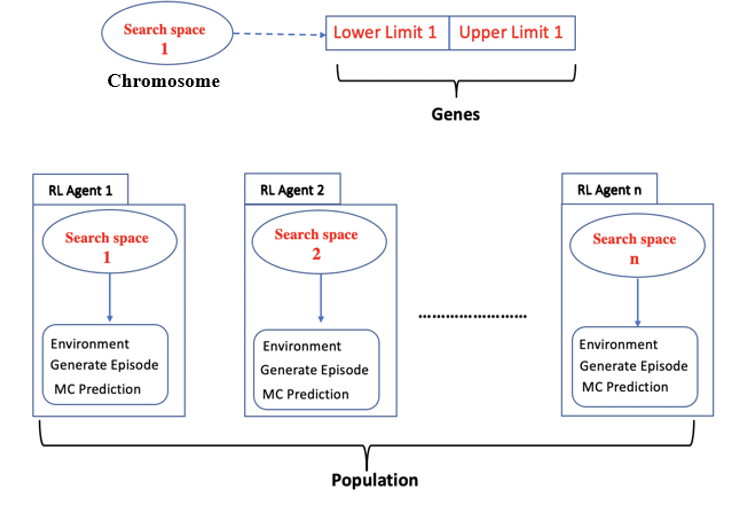}
      \caption{Genetic Algorithm elements in our GA-RL formulation for MTD.}
      \label{fig:GA-RL}
\end{figure}

Figure \ref{fig:GA-RL} illustrates how the Genetic Algorithm's (GA) components are represented to be integrated into the RL-MTD(RL) framework. In this model, we have:
\begin{itemize}
    \item \textbf{Gene:} It's an initial random \textit{lim} int value used to define the Lower Limit (LL) and Upper Limit (UL) of a search space corresponding to the datatype of setting(P) as mentioned in section 3. 
    \item \textbf{Chromosome:} It's an individual search space whose range is composed be gene values.
    \item \textbf{Population:} It is a pool made up of different RL-MTD agents each characterized by its search space range. (where each RL agent acts as a chromosome)
\end{itemize}

The objective is to evolve these different search spaces through the GA with multiple generations. Each RL agent draws from its unique settings search space range to generate random initial configuration states, which are then utilized by the RL algorithm. This algorithm encompasses environment setup, episode generation, and Monte Carlo (MC) prediction functionalities. The initial population consists of a variety of RL agents. Through the iterative processes of Selection, Crossover, and Mutation, this population undergoes evolution across generations. At the culmination of these generations, we identify and select the most effective RL agent from this pool. The selection criteria focus on the agent that achieves the maximum rewards or creates the most secure configurations.

A step-by-step construction of incorporating GA into our RL -MTD is mentioned below. We follow the logic for Selection and Crossover steps similar to the neuroevolution algorithm described in \cite{blog}:
\begin{enumerate}
    \item Treat the \textit{search space} the agent draws from as a chromosome.(We will address the agent's search space as an entire agent for ease of use and consistency throughout the paper)
    \item Agent's parameter,  $lim$ of search range will act as its genes.
    \item The fitness score of secure configurations generated will act as the chromosome’s fitness (i.e. higher the fitness score, the higher the likelihood of survival).
    \item The first iteration starts with $n$ number of agents (search range), all with randomly initialized parameters.
    \item  \textbf{Selection:} By pure chance, some of them will perform better than others. The survival of the fittest option is then implemented by simply excluding the weakest agents from consideration and considering a certain percentage of agents.
    \item \textbf{Crossover:} It is quite risky to swap parameters/genes for the simple reason that it might disturb the best-performing agents' search space range limits. Hence, we rather replicate the selected agents for the next iteration until we reach $n$ agents again for the next iteration.  
    \item \textbf{Mutation:} We modify agents produced during the crossover step, by adding or subtracting a small noise (value) to its parameter(limit). This step ensures we get to explore the neighborhood around the parameters of the best agents in the next iteration.
    \item To secure the best agents from a probable reduction in performance due to the mutation step, we decided to keep the top-performing agent as is (without adding noise).
\end{enumerate} 

\subsubsection{PSO-RL}
\begin{figure}[h]
 \begin{center}
      \includegraphics[width=0.9\linewidth]{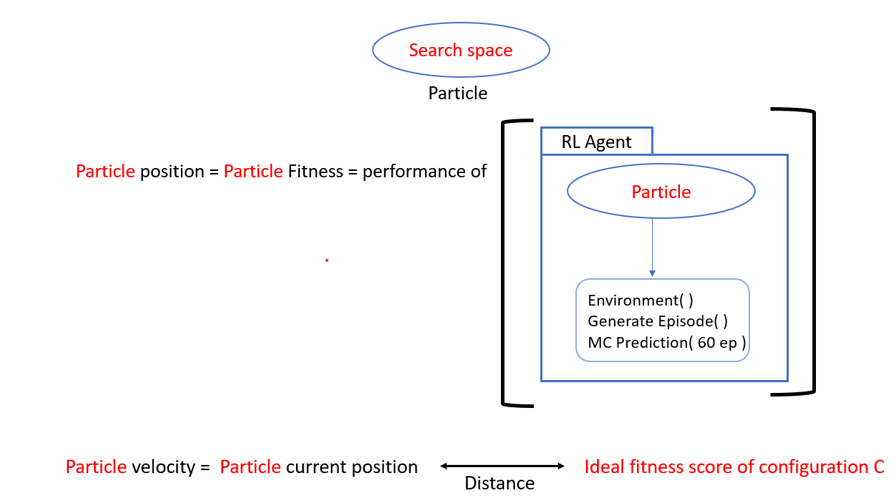}
      \caption{PSO elements in our PSO-RL formulation for MTD.}
      \label{fig:ps-RL}
    \end{center}
\end{figure}

The components in our PSO-RL model as shown in Figure \ref{fig:ps-RL}  are as follows: 
\begin{itemize}
    \item \textbf{Particles / Swarm:} In analogy to Figure \ref{fig:GA-RL}, a particle corresponds to a chromosome in GA, representing a limit \textit{lim}  used to define the search space range. Similarly, akin to the population in GA, a swarm, comprised of a collection of particles, represents a list of limit values defining various search space ranges, thus forming a swarm of distinct RL agents.
    \item \textbf{Particle Position:} Particle represents an individual search space. We denote the particle's position in terms of its fitness. Here fitness is measured by the performance of the RL-MTD model when supplied with that particle.\newline
    $Fitness(particle) = performance(RL-MTD[particle)]$ \newline
    where performance is the agent’s average performance (fitness score) over 60 episodes.
    \item \textbf{Global Best:} In its simplest form, global best refers to the best value of a fitness score among a set of particles. Each particle updates its position at the final stage of the search space exploration, with the best position being identified as the global best.
     \item \textbf{Particle Velocity:} This represents how far a particle position is from the ideal position (fitness). This indicates the moving rates for each particle within the search space. Our study takes advantage of this part to determine the distance between the best particle position and the ideal secure position.

\end{itemize}

Just like, GA, the goal here is to find the most optimal particle (search space) for our RL-MTD algorithm. The PSO-RL algorithm uses the particle's position, velocity, and global best and is run for 100 generations as per the standard PSO flowchart shown in Figure \ref{fig:PSOflow} to find the most optimal search space for the RL-MTD  model to generate more secure configurations.

\subsection{Bio-inspired MTD Algorithm Design}
In this section, we describe the E-RL algorithms for finding the best-performing agent(search space) to generate a more secure configuration.
\subsubsection{GA-RL}
Algorithm \ref{alg:E-RL} shows how the GA-RL was adapted for finding the best RL agents in generating secure configuration.
\footnote{\raggedright\url{https://github.com/paraschopra/deepneuroevolution/blob/master/openai-gym-cartpole-neuroevolution.ipynb}\par}

\begin{algorithm}[!h]
    \caption{GA\_RL Algorithm to find best RL agents}\label{alg:E-RL}
    \begin{algorithmic}[1]
        \Procedure{perform\_GA}{}
            \State $num\_agents \gets $ n  \Comment{Initialize n number of agents}
            \State $agents  \gets $ generate\_random\_agents(num\_agents)
            \State $top\_limit \gets k$ \Comment{\# of top agents to consider as parents}
            \For {$gen$ in $range(X)$} \Comment{run evolution until X generations}
                \State $rewards \gets $ run\_agent(agents) \Comment{return rewards of agents}
                \State $selected\_agents \gets $ Select agents of top $k$ rewards \Comment{Selection}
                \State $children\_agents \gets $ Randomly choose $k$-1 agents from $selected\_agents$    \Comment{Crossover}
                \State $mutated\_agents \gets $ Mutate($children\_agents$) \Comment{children agents after Mutation}
                \State $agents \gets mutated\_agents$ \Comment{kill all agents, and replace them with mutated children}
            \EndFor
            \State $Best\_agent \gets$ Select agent from $agents$ with maximum reward
        \EndProcedure
        \State {\bf return} $Best\_agent$
    \end{algorithmic}
\end{algorithm}

Algorithm \ref{alg:E-RL}  follows the pseudo-code mentioned in the previous section. 
\begin{itemize}
    \item With population size set to $n$ ($num\_agents$), we generate agents randomly in the first iteration using the function\newline
    $generate\_random\_agents$ described in Algorithm \ref{alg:gen_agent}
    
    \item We set the maximum number of generations to run the loop to $X$.
    \item The $run\_agent$ function is used in each generation to run all randomly generated agents and get their performance (mean fitness score).(Algorithm \ref{alg:run_agent})
    \item \textit{Selection} : Out of $n$, select only top $k$ as parents ($top\_limit$) where $k<n$.
    
    \item \textit{Crossover}: As mentioned before, we replicate the selected agents instead of swapping parameters. Among top $k$ parent agents,  $k$-1 agents are randomly chosen to make children for the next iteration.
    \item \textit{Mutation}: In the $mutate$ function, we add or subtract a small noise(value) to the parameter if the value of the agent is greater than a random number. (Algorithm \ref{alg:mutate})
    \item After we have child agents as parents, we iterate over the loop again until all generations are done or we find a good performing agent. 
\end{itemize}

\begin{algorithm}[!h]
    \caption{GENERATE AGENTS}\label{alg:gen_agent}
    \begin{algorithmic}[1]
        \Procedure{generate\_random\_agents}{$num\_agents$}
            \State $agents \gets [ ] $ 
            \For {$1$ to $num\_agents$ }
            \State $lim \gets $ Pick \ random \ number from\ range (1, N)  \Comment{N is an integer}
            \State $agents.append(lim)$
            \EndFor
        \EndProcedure
        \State \textbf{return}  $agents$
    \end{algorithmic}
\end{algorithm}

In Algorithm \ref{alg:gen_agent}, function $generate\_random\_agents$ is used to generate $num\_agents$ number of agents . Here the agent represents limit value $lim$ which is an integer value randomly drawn  between 1 and $N$ (hyperparameter). This $lim$  decides the search space range from which the agent picks from.

\begin{algorithm}[!h]
    \caption{RUN AGENT}\label{alg:run_agent}
    \begin{algorithmic}[1]
        \Procedure{run\_agent}{$agents$}
            \State $reward\_agents \gets [ ] $ 
            \For {$ag$ in $agents$ }
            \State $rwrd = run\_RL\_MTD(ag)$ \Comment{Call the $run\_RL\_mc$ func which takes as i/p the agent(i.e lim) }
            \State $reward\_agents.append(rwrd)$
            \EndFor
        \EndProcedure
        \State \textbf{return}  $reward\_agents$
    \end{algorithmic}
\end{algorithm}

In Algorithm \ref{alg:run_agent}, the procedure $run\_agent$ takes a list of agents  $agents$ as input.  For each agent $ag$ in $agents$ list,  $run\_RL\_MTD()$ (Figure 5) is called which takes the agent as input and runs the base RL-MTD model with the new search space ($ag$). The MC prediction (Algorithm \ref{alg:MCP}) method in the RL-MTD model returns the agent's average performance (fitness score) over 60 episodes which is stored in $rwrd$.

\begin{algorithm}[!h]
    \caption{MUTATION}\label{alg:mutate}
    \begin{algorithmic}[1]
        \Procedure{mutate}{$children\_agent$}
            \State $mutate\_agent \gets children\_agent $ 
             \If {$child\_agent$ $>$ $random.random()$}  
                \State $mutate\_agent -= noise$  \Comment{hyperparameter}
            \Else
                \State $mutate\_agent += noise$ 
             \EndIf
        \EndProcedure
        \State \textbf{return}  $mutate\_agent$
    \end{algorithmic}
\end{algorithm}

\subsubsection{PSO-RL}

In Algorithm \ref{alg:initialpso}, we initialize all the N particles in a swarm with random integers. Each particle's starting position will have the same random integer and the same goes for the particle's velocity. The optimal difference is the hyperparameter we experiment with which indicates the threshold value. This threshold value is to check how far is the particle search space from the optimal one in terms of fitness scores generated from their corresponding RL-MTD agents.

\begin{algorithm}[!h]
    \caption{INITIALIZING N PARTICLES}\label{alg:initialpso}
    \begin{algorithmic}[1]
     \Procedure{initialize}{}
            \State $swarm \gets $ Pick \ N random\ integers \Comment{This is the limit values representing each particle }
            \For {$particle$ in $swarm$}
                \State $particle\_position \gets $ Start \ with \ random \ integer 
                \State $particle\_velocity \gets $  Start \ with \ random \ integer
            \EndFor
            \State $global\_best \gets 0 $ 
            \State $generations \gets 100$ 
            \State $optimal\_difference \gets d$
            \State $ideal\_fitness \gets $ Fitness\ Score\ Ideal\ C \Comment{Its the score of fully secure config}
      \EndProcedure
    \end{algorithmic}
\end{algorithm}

The primary goal of Algorithm \ref{alg:particleposition} is to calculate each particle's current position and update the $gbest$ with the maximum value. The procedure $run\_RL\_MTD()$ (base MTD-RL algorithm)) is called for each value in the swarm. 
The obtained value is then contrasted with the preceding value. The global best receives the new value if the outcomes attained are the highest value.

\begin{algorithm}[!h]
    \caption{PARTICLE POSITION}\label{alg:particleposition}
    \begin{algorithmic}[1]
        \Procedure{maximum\_particle\_position}{}
            \For {$particle$ in $swarm$}
            \State $new\_pp = run\_RL\_MTD(particle)$ \Comment{Call the $run\_RL\_mc$ func }
                \If {$new\_pp$ $>$ $particle\_position[particle]$}  
                    \State $particle\_position[particle] = new\_pp$
                \EndIf
                \State $global\_best = max(new\_pp)$ 
                \EndFor
        \EndProcedure
        \State \textbf{return}  $global\_best$
    \end{algorithmic}
\end{algorithm}

The distance between the ideal fitness and the particle's current position(search space's fitness) is represented by particle velocity. The lesser the distance, the more optimal the particle as it makes the base RL-MTD model generate an almost secure configuration. In Algorithm \ref{alg:particlevelocity}, we are trying to calculate the current distance of the particle (line 3) and we update the particle's old velocity with the current one if the latter is less than the former even though it's still greater than optimal difference (line 4,5). This means it's slowly approaching optimal difference.

\begin{algorithm}[!h]
    \caption{PARTICLE VELOCITY}\label{alg:particlevelocity}
    \begin{algorithmic}[1]
        \Procedure{minimum\_particle\_velocity}{}
            \For {$particle$ in $swarm$}
            \State $new\_pv = ideal\_fitness - particle\_position[particle] $
                \If {$(new\_pv$ $>$ $optimal\_difference)$ and $new\_pv$ $<$ $ particle\_velocity[particle]) $}  
                    \State $particle\_velocity[particle] = new\_pv$
                \EndIf
                \EndFor
        \EndProcedure
         \State \textbf{return}  $new\_pv$
    \end{algorithmic}
\end{algorithm}

In Algorithm \ref{alg:socialinfluence}, $social\_influence$ seeks to improve the particle value in regards to the intermediate best particle via social influence in a particular generation. It attempts to deduct social influence value)or add it depending on whether the current particle is greater or less than the intermediate best particle, thereby trying to converge all the particles to the near-optimal particle.

\begin{algorithm}[!h]
    \caption{SOCIAL INFLUENCE}\label{alg:socialinfluence}
    \begin{algorithmic}[1]
        \Procedure{social\_influence}{} \Comment{particle value learning to be closer to the best one via social influence}
          \For {$particle$ in $swarm$}
             \If {$particle $ $>$ $best\_particle$}  
                \State $particle] -= influence$  \Comment{random float social influence value}
            \Else
                \State $particle] += influence$  
             \EndIf
         \EndFor
        \EndProcedure
    \end{algorithmic}
\end{algorithm}

To determine the optimal $particle$ (search space), Algorithm \ref{alg:generationpso} which is the PSO algorithm is run over 100 generations  
\begin{algorithm}[!h]
    \caption{RUN PSO GENERATIONS}\label{alg:generationpso}
    \begin{algorithmic}[1]
            \For {$gen$ in $generation$ } \Comment{ 100 iteration for finding the best limit\_value}
            \State $maximum\_particle\_position()$
            \State $minimum\_particle\_velocity()$
            \State $social\_influence()$
                \EndFor
            \State \textbf{return}  $particle$
    \end{algorithmic}
\end{algorithm}

%% file: 6ExperimentsandResults.tex
\section{Experiments and Results}
\label{sec:eval}
In this section, we elucidate the experimental setup for each algorithm used in implementing MTD: RL-MTD, GA-RL and PSO-RL where we ran each of them on 4 SUT case studies and compared their performance results. We intend to compare each of these models to determine the most effective approach to generating secure configuration using the MTD approach. 

\subsection{Experiment Set Up}
\subsubsection{RL-MTD}
We implemented the algorithm using python 3.6 with libraries numpy and pandas. After much experimentation, we set the following hyperparameters:
\begin{itemize}
    \item lim  = 10 (section 3)
    \item parameter score = HIGH(secure): 800; LOW(not secure): 8 (Section 2.2.3)
    \item Threshold = any value in \texttt{range}:\\
    $(\texttt{max\_score} - \texttt{val}, \texttt{max\_score})$,
    where \texttt{val} is a hyperparameter set to 800, 
    and \texttt{max\_score} is the ideal total fitness score of configuration C  when all parameters are securely set (goal state).  In the action policy, we discourage the agent from selecting the "change" action  if \texttt{fit(C)} falls within this threshold range, as it indicates that all parameters of C are securely set except one. This ensures that the likelihood of the agent choosing action 0 remains high (0.8) in such cases.

\end{itemize}

\subsubsection{GA-RL}
We implemented GA-RL algorithm using Python 3.6 with libraries numpy and pandas. After much experimentation, we set the following hyperparameters:
\begin{itemize}
    \item number of agents n  = 25
    \item number of top agents k = 5
    \item number of generations X = 100
    \item N = 25
    \item noise = 0.05
\end{itemize}
\textbf{Fitness function for chromosome:} 
To calculate rewards for different RL agents, we used the complete RL-MTD algorithm as the fitness function. It returns the average of the scores generated by the Monte Carlo (MC) prediction function,  evaluated over episode counts of 20 and 60.

\subsubsection{PSO-RL}
We implemented the PSO-RL algorithm using Python~3.6, with the libraries \texttt{numpy} and \texttt{pandas}. We set the hyperparameters as follows:
\begin{itemize}
    \item Swarm size  = 30
    \item Particle values (lim): integers = [1....30]
    \item Initial Particle position = 0
    \item Idea Fitness = Maximum fitness score of the most secure configuration based on the selected SUT.
    \item number of generations  = 100
    \item influence = 0.05
\end{itemize}
Given that configuration C varies across different SUTs due to its diverse parameters, it became evident that employing identical optimal difference and initial particle velocity values to be implemented for all case studies was inappropriate. This is because the values depended on the specific parameters and their quantity within each SUT's configuration. Consequently, we undertook an exploration of various optimal particle and initial particle velocity values to identify the most effective combinations tailored to each SUT. The ultimate goal is to recognize the optimum balance between Particle Velocity and the Optimal Difference. This leads to the attainment of maximum or highly improved results within the PSO-RL framework. 

We elaborate on the ideal set of hyper-parameters examined for PSO, which comprises the Optimal Difference and Particle Velocity for each case study. 
The ideal set refers to the optimal combination of parameters that led to the most secure and effective results in our experiments.
Considering these two hyper-parameter values as a tuple (Optimal Difference, Particle Velocity), our best combinations for each case study are as follows:

\begin{itemize}
\item Window 10: (20, 300)
\item McAfee: (300, 500)
\item Microsoft Excel 2016: (160, 200)
\item Microsoft Office 2007: (120, 1000)
\end{itemize}


\subsection{Results on case studies}
This section demonstrates the performance of all three models: RL-MTD, GA-RL, PSO-RL,  in generating secure configurations and reports the results. Once we get the optimal search space range for both GA-RL and PSO-RL, we run the base RL-MTD with those optimized search spaces and capture their results. We then compare their performance with the base RL-MTD with no optimization.
We selected 4 SUT case studies and the corresponding parameters from the STIG website namely:
 \begin{enumerate}
     \item Windows 10 (59 parameters)
     \item McAfee (14  parameters)
    \item MS Excel (20  parameters)
    \item MS Office (21  parameters)
 \end{enumerate}
These SUTs contain a good number of configuration parameters whose domains are diverse enough. \newline
We executed our developed scripts for varying numbers of episodes and captured the best fitness scores  (i.e., the value of state~$V$, which indicates how secure a given state is).  Figures~\ref{fig:performance} illustrate the trend of fitness scores obtained through the episodes where the x-axis is the episode counts (i.e., between 20-500 episodes); whereas, the y-axis holds the normalized values of fitness scores. More specifically, the normalized fitness score value of 0.0 represents the least secure attained by the agent; whereas, the value 1.0 is the most secure fitness score. The normalization on fitness scores are performed as follows:

\[
normalized(fs) = \frac{fs - min(fs)}{max(fs) - min(fs)}
\]


where $min(fs)$ for a given fitness score $fs$ is the minimum fitness score of the configuration which is equal to the total sum of the fitness scores for all parameters when they are all set to LOW. Similarly, $min(fs)$ for a given fitness score $fs$ is the maximum fitness score of the configuration which is equal to the total sum of the fitness scores for all parameters when they are all set of HIGH. More specifically,:

\[
min(fs) = \sum_{S_{i} = 1}^{n} LOW\ \ \  ;\ \ \  max(fs) = \sum_{S_{i} = 1}^{n} HIGH 
\]



where $S_{i}$ is the $i$th parameter and $n$ is the total number of parameters.


We now analyze the results for each case study which is also summarized in Table 1.
\subsubsection{Window 10 Case Study}
Upon examining Figure \ref{fig:Hwindows10}, an initial observation reveals that RL-MTD demonstrates the least security performance. However, an instant competition unfolds between GA-RL and PSO-RL (labeled ERL). 
Ultimately, it becomes evident that in this context, PSO-RL outperforms GA-RL, although a slight difference can be depicted. 
Moreover, in this case study, PSO-RL outperformed 17 times in comparison to GA-RL.

\subsubsection{McAfee Case Study}
In this case study, substantial fluctuations are observed across all experiments; however, RL-MTD consistently exhibits inadequate performance compared to the other two methods. Upon initial glance of GA-RL and PSO-RL (labeled ERL) in Figure \ref{fig:Hmsmcoffe}, notable oscillations are evident. Furthermore, both PSO and GA equally performed better than base model for a count of 8 episodes. A distinct pattern emerges in these two methods, while they promote similar performance. 
Specifically, in corresponding episodes, both methods experience periods of insecurity, followed by episodes where performance improves. 
For instance, in episode number 40, the performance of neither algorithm is remarkable, but in the subsequent episode, a notable improvement can be illustrated.

\subsubsection{Microsoft Excel 2016 Case Study}

In this case study, illustrated in Figure \ref{fig:HMSExcel}, RL-MTD and GA-RL (labeled ERL) display predominantly similar and fluctuating behaviors. 
The frequency of episodes where each outperforms the other is comparable.
Although RL-MTD may exhibit better performance in specific episodes, the reverse occurs for GA. 
Despite the oscillations perceived in PSO-RL as a third graph, it consistently surpasses the other two methods in most instances. 
Moving to the episode counts analysis, it becomes evident that the results are outstanding when employing the PSO-RL model in the Microsoft Excel 2016 case study.
In the majority of episodes, PSO-RL demonstrates upper-level performance compared to other methods.

\subsubsection{Microsoft Office 2007 Case Study}

\begin{figure*}[h]
\centering
    \begin{subfigure}[b]{0.45\textwidth}
      \includegraphics[width=\textwidth]{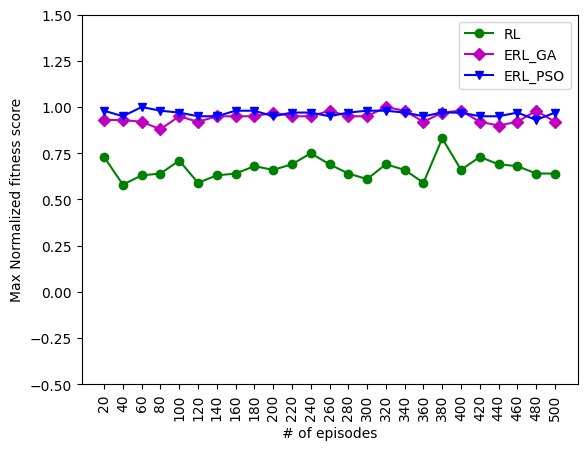}
      \caption{Windows10}\vspace*{0.1in}
      \label{fig:Hwindows10}
    \end{subfigure}%
    \begin{subfigure}[b]{0.45\textwidth}
      \includegraphics[width=\textwidth]{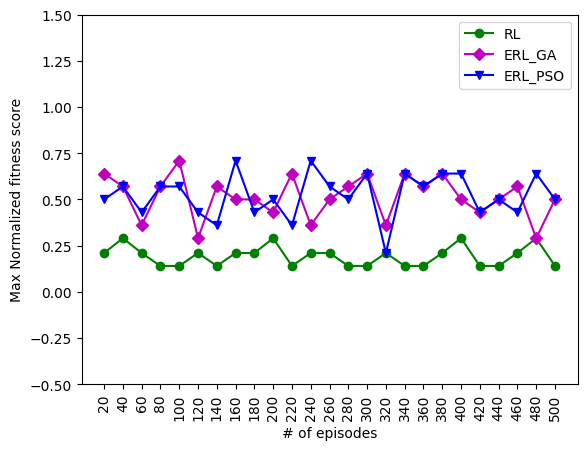}
      \caption{McAfee}\vspace*{0.1in}
      \label{fig:Hmsmcoffe}
    \end{subfigure}%
    
    \begin{subfigure}[b]{0.45\textwidth}
      \includegraphics[width=\textwidth]{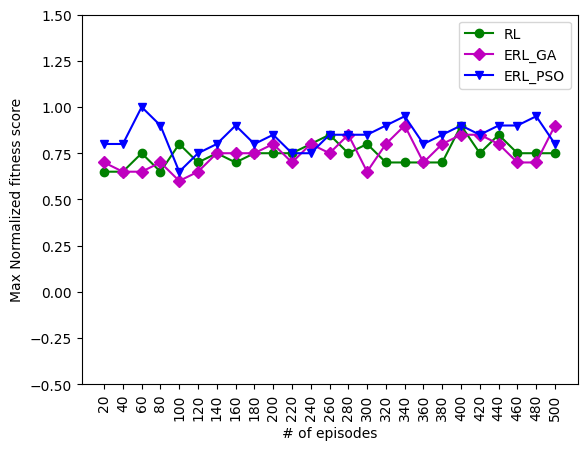}
      \caption{MSExcel}\vspace*{0.1in}
      \label{fig:HMSExcel}
    \end{subfigure}%
    \begin{subfigure}[b]{0.45\textwidth}
      \includegraphics[width=\textwidth]{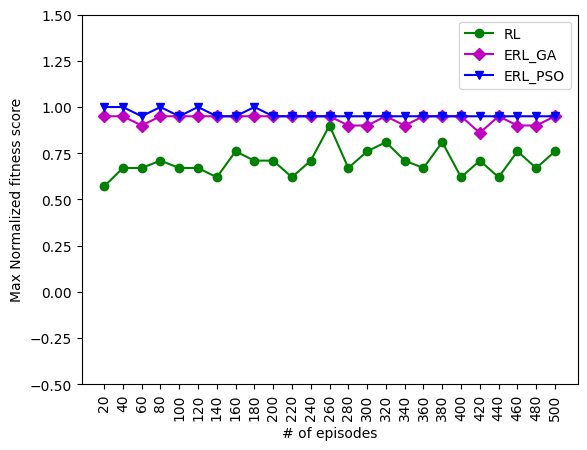}
      \caption{MSOffice}
      \label{fig:Hmsoffice}
    \end{subfigure}%
    \vspace*{0.1in}
    \caption{Performance comparison of the 3 models: MTD-RL(labeled RL), GA-RL (ERL-GA), and PSO-RL (ERL-PSO) in generating secure configurations for 4 case studies. The x-axis shows the number of episodes each of the models were trained on and the y-axis shows the normalized fitness each of these models was able to achieve. 1 being the max fitness score: (all parameters securely set) and 0 being the most insecure. From the trend, we see both bio-inspired algorithms outperformed the base RL-MTD model for most of the SUTs reaching almost the most secure config across all episodes. However,  between GA-Rl and PSO-RL, its a close call but as per our analysis in Table 1, PSO-RL was the best performing model by a margin. }
    \label{fig:performance}
\end{figure*}

As depicted in Figure \ref{fig:Hmsoffice}, a clear distinction is evident in the performance between RL-MTD and Bio-inspired methods (labeled ERL), including GA-RL and PSO-RL, upon initial glance. 
Although RL-MTD attempts to reach Bio-inspired methods, within one episode, there is a substantial difference in subsequent episodes. 
A slight discrepancy is observed by moving to the GA-RL and PSO-RL methods. 
In most episodes, GA-RL and PSO-RL demonstrate approximate performance. 
However, GA-RL slightly exhibits less security in a few instances. 
It is evident that Bio-inspired methods which are GA-RL and PSO-RL, achieve the highest level of security performance in almost all episodes iteratively. 
This underlines that both GA-RL and PSO-RL attain the higher-rank secure performance.





\begin{table}[h!]
\resizebox{0.47\textwidth}{!}{%
\begin{tabular}{|c|c|c|}
\hline
 Case Studies & Count of episodes (RL, GA-RL, PSO-RL) & Best Algorithm \\
 \hline\hline
 Window 10 & (0, 6, \textbf{17}) & PSO-RL\\ \hline
 McAfee & (0, \textbf{8, 8}) & GA-RL , PSO-RL\\\hline
 MS Excel 2016 &  (0, 2, \textbf{21}) & PSO-RL\\ \hline
 MS Office 2007 &  (0, 0, \textbf{12}) & PSO-RL\\ \hline
 \hline
\end{tabular}
}%
\vspace*{0.2in}
\caption{This table summarizes the results/analysis (Figure \ref{fig:performance}) of the 3 models on each SUT and shows PSO-RL was the clear winner among all 3. We considered 25 number of episodes for each model namely, RL-MTD, GA-RL, and PSO-RL. The values in tuples illustrate the total count of episodes in which each model outperformed the rest two. For instance, for the Windows 10 case study, PSO-RL outperformed others 17 times, while GA-RL did better 6 times. Having said that, in contrast to PSO-RL and GA-RL, base RL-MTD did not achieve better in generating configuration far nay SU. When the cumulative number of episodes is less than 25, the rest of the episodes show equal values between the methods. }
\label{table:psoresults}
\end{table}

\vspace{-1.2em}

%% file: 7Discussion.tex
\section{Discussion}

As seen in the results, the base RL-MTD approach exhibits less secure performance across all case studies.  The incorporation of the optimal search space derived from both GA and PSO significantly enhanced the performance of our base RL-MTD model compared to its performance without the optimal search space. Interestingly, our analysis revealed that while there was a notable performance improvement when utilizing the optimal search space from either GA or PSO, the difference in performance between GA and PSO was marginal. This suggests that both GA and PSO were effective in searching for an optimal search space to enhance the base model's performance. However, \textbf{the clear winner was PSO-RL}. Our findings underscore the efficacy of both optimization techniques in facilitating the identification of an optimal search space conducive to generating more secure configurations within the RL framework.
In some illustrations, such as Windows 10 and Microsoft Office 2007, the minor improvements prevailed. Nevertheless, PSO ultimately proves better results in other cases like Microsoft Excel 2016, despite slight fluctuation. 
It should be noted that in certain case studies, such as McAfee, significant performance fluctuations lead to equal outcomes for GA and PSO. 
A key point to consider is that in each case study, optimization drives experimentation to enhance outcomes and performance.\newline
\textbf{NOTE: As previously noted, the application of MTD to address security challenges in misconfigured software is relatively novel, resulting in the absence of an established benchmark for comparison in existing literature. This study aims to be viewed as the construction of a proof-of-concept (POC), showcasing the potential of an MTD defense strategy from the defender's standpoint in generating dynamic secure configuration. It is crucial to acknowledge that further research and refinement are imperative to bolster this strategy and guarantee its overall effectiveness.}

%% file: 9LiteratureReview.tex
\section{Related Work}
\label{sec:review}


Examining the relevant literature closely reveals that there hasn't been adequate discussion of the intended motivation.
By scrutinizing the related work can be noticed that the targeted motivation has not been addressed much.
Gao et al.~\cite{gao2021reinforcement} countered the moving target defense method by implementing Reinforcement Learning to address the DDoS attack. 
The findings of their experiments advocate that the reinforcement learning adjustment on MTD affects the results more successfully, and enhances the algorithm performance.
Zhang et al.~\cite{zhang2023disturb} generated the Moving target defense method draws on deep reinforcement learning to protect from cyber security attacks. The Markov decision process (MDP)model was used to design the MTD technique to train on scanning behavior. The analysis of the developed model revealed that the scanning time was diminished effectively. 
Eghtesad et al.~\cite{eghtesad2019deep} applied the reinforcement learning technique to ameliorate the MTD method performance in their study. 
The authors evaluated the numerical results demonstrating the performance of the trained model. The outcomes depict that the developed model has the exceptional ability to recognize the optimal policies in a defined environment.
Li et al. ~\cite{li2023optimal} utilized the MTD technique to mitigate the potential threats and attacks that might be encountered in the container cloud environment.
The proposed model exploits the advantage of deep reinforcement learning along with the Markov decision process in the optimized MTD. 
The model demonstrates the improvement of the efficacy of defense significantly.
The authors of~\cite{john2014evolutionary} implemented the Genetic Algorithm in Moving Target Defense to find the optimized secure configuration. 
John et al. stated that the variety of configurations is potentially improved over time regarding the environment. 
The experiment findings confirm that the GA method is sufficiently capable of identifying the optimal secure configuration. Zhang, et al.~\cite{electronics13040734} introduced the Moving Target Three-way Evolutionary Game Defense Model(MTTEGDM) in network security. 
The model, which concentrates on offering adjustable defense decisions, combines evolutionary games and signal games. 
The authors employed the action and rewards strategy to optimize the defense. 
Findings and analyses reveal that the Monte Carlo simulation performs better compared to former designs. 
However, defending from various attack behaviors is still ongoing.

%% file: 8ConclusionandFuturework.tex
\section{Conclusion and Future work}
\label{sec:conclusion}
In conclusion, our in-depth research tackles the pervasive challenge of security misconfiguration within software systems—a problem that leaves systems open to exploitation. By integrating bio-inspired algorithms, specifically the Genetic Algorithm (GA) and Particle Swarm Optimization (PSO), into our previously established RL-MTD model, we've advanced the model's proficiency in continuously generating secure and dynamic configurations. Through rigorous comparative analysis, we've ascertained that both the GA-RL and PSO-RL enhancements not only refine the search space for potential configurations but also surpass the original RL-MTD framework in their ability to produce robust configurations against an array of software systems put to the test (SUTs). Noteworthy is the slightly superior performance of PSO-RL in the majority of these scenarios, making it the best-performing model due to a nuanced edge in its search strategy. Our findings represent a significant contribution to the domain of proactive cybersecurity measures. By employing Moving Target Defense (MTD) enriched with machine learning and bio-inspired algorithms, we present an innovative and efficacious strategy to proactively MTD defense strategy from a defender perspective, thereby reinforcing the security posture of software configurations against the dynamic landscape of cyber threats.\newline
For future work, we envision expanding our research to transform the current game model into a more complex and realistic scenario. The next step is to develop a two-player version of the game, where one player is the defender, maintaining secure configurations, while the other acts as an attacker, probing and exploiting vulnerabilities. Additionally, since software applications do not operate in isolation but interact with each other, it is crucial to examine these interactions and how they may impact the Moving Target Defense (MTD) strategy. We also need to study how different configuration parameters influence each other. Ultimately, this research is a stepping stone towards a deeper understanding of how to protect software from misconfiguration threats using MTD, aiming to build a more resilient digital infrastructure.
